\documentclass[%
 aip,
 amsmath,amssymb,
 reprint,%
]{revtex4-1}
\usepackage{graphicx}% Include figure files
\usepackage{dcolumn}% Align table columns on decimal point
\usepackage{bm}% bold math
\usepackage[utf8]{inputenc}
\usepackage[T1]{fontenc}
\usepackage{etoolbox}
\usepackage[dvipsnames]{xcolor}
\usepackage{amsmath}
\usepackage{txfonts} % diamondblack shape
\usepackage{soul} %strikeline

\makeatletter
\def\@email#1#2{%
 \endgroup
 \patchcmd{\titleblock@produce}
  {\frontmatter@RRAPformat}
  {\frontmatter@RRAPformat{\produce@RRAP{*#1\href{mailto:#2}{#2}}}\frontmatter@RRAPformat}
  {}{}
}%
\makeatother
\begin{document}

\preprint{AIP/123-QED}

\title{Diagonal Born–Oppenheimer Corrections in Condensed-Phase Ring Polymer Surface Hopping}

\author{Dil K. Limbu}
\author{Sandip Bhusal}
\author{Diana M. Casta\~{n}eda-Bagatella}
\author{Farnaz A. Shakib}
 \email{shakib@njit.edu}
\affiliation{ 
Department of Chemistry and Environmental Science, New Jersey Institute of Technology, Newark, New Jersey 07102, United States
}

\date{\today}% It is always \today, today,

\begin{abstract}
Ring polymer surface hopping (RPSH) is a  mixed quantum-classical dynamics method for incorporating nuclear quantum effects (NQEs) into nonadiabatic dynamics simulations via the extended phase-space of a classical ring polymer. Here, we systematically investigate several variants of RPSH in the frameworks of centroid and bead approximations (RPSH-CA and RPSH-BA) in modeling the dynamics of the spin-boson system across different reaction regimes, reorganization energies, and temperatures. Moreover, the effects of including the diagonal Born–Oppenheimer correction (DBOC) on the performance of the RPSH-CA and RPSH-BA methods are investigated. Our simulations of symmetric potentials, i.e., without energy bias, show that the RPSH-CA method, where nonadiabatic transitions are handled at the centroid level, is satisfactorily accurate and robust across different reaction regimes. Adding DBOC improves the method's accuracy in specific intermediate and nonadiabatic reaction regimes at low temperature. Overall, the effect of DBOC in RPSH-CA is in moderation compared to conventional fewest-switches surface hopping method where DBOC over-damps the dynamics significantly and reduces accuracy considerably, especially at low temperatures. However, the RPSH-CA and its DBOC variant struggle in simulations of asymmetric potentials specially at low temperatures. On the other hand, RPSH-BA results, where nonadiabatic transitions are handled at the level of individual beads of the ring polymers, are generally unreliable unless in the high temperature adiabatic reaction regimes with symmetric potentials. The inclusion of DBOC is not particularly helpful in remedying this erratic behavior. Our findings clarify when geometric corrections are beneficial or detrimental to nonadiabatic simulations using RPSH, providing practical guidance for atomistic condensed-phase applications.

\end{abstract}

\maketitle

\section{\label{sec:level1}Introduction}
Path integral formalism of quantum mechanics\cite{Feynman:1948,Feynman:2010} achieves the preservation of the quantum Boltzmann distribution via an isomorphic classical Hamiltonian.\cite{Parinello:1984} This feat is highlighted in the development of approximate methods, centroid molecular dynamics (CMD)\cite{Cao_II:1994,Cao_III:1994,Cao_IV:1994,augJang1999} and ring polymer MD (RPMD)\cite{Craig:2004,Craig:2005a,Craig:2005}, for investigating chemical dynamics on a single electronic surface.\cite{Cao:1996,Geva:2001,Collepardo-Guevara:2008,Habershon:2013} Nuclear quantum effects (NQEs) such as zero-point energy and tunneling are included in adiabatic dynamics simulations via the extended phase-space of a ring polymer created from \textit{n} replicas of the physical system, also known as beads, coupled by harmonic forces. While RPMD and CMD are not derived from first principles, they can be recovered from Matsubara dynamics with a generalized Kubo-transformed time correlation function formalism,\cite{Hele:2015} putting them on a strong ground for further developments beyond the adiabatic reaction regime.\cite{Markland:2018,Ananth:2022}

RPMD formalism has been successfully utilized in the description of electron transfer\cite{Menzeleev:2011} and proton-coupled electron transfer\cite{Kretchmer:2013} reactions. Despite its success, the computational cost associated with quantizing both nuclear and electronic degrees of freedom by ring polymers, with more than 1000 beads needed for the latter,{\textcolor{blue}{\cite{Menzeleev:2011,Kretchmer:2013}}} renders this treatment expensive for multi-electron/multi-proton nonadiabatic dynamics simulations. Still, inclusion of NQEs via the extended phase-space of a ring polymer is more affordable than other sophisticated methods, such as the mixed quantum classical Liouville approach\cite{Kapral:1999,Hanna:2005,Brown:2021} where both proton and electron wavefunctions need to be expanded in terms of harmonic basis sets on top of the millions of trajectories needed to be run for convergence.\cite{Shakib:2014,Shakib:2016a,Shakib:2016b} Among multi-state RPMD formulations, ring polymer surface hopping (RPSH)\cite{Shushkov:2012,Shakib:2017} stands out for its simplicity, ease of implementation, and inherent capability of the method for including NQEs into nonadiabatic MD simulations. Describing the nonadiabatic electronic transitions according to the fewest switched surface hopping (FSSH)\cite{Tully:1971,Tully:1990,Hammes-Schiffer:1994} algorithm, RPSH is conveniently suitable for multi-electron nonadiabatic dynamics simulations. In a different class than nonadiabatic instanton theory\cite{Schwieters:1998,Schwieters:1999,Jang:2001} and mean-field RPMD,\cite{Ehrenfest:1927,Duke:2016} it does not rely on a thermally averaged description of the electronic nonadiabatic dynamics but follows the dynamics explicitly on each adiabatic surface, making it more suitable for simulating scattering events and calculating branching probabilities. Finally, as in the original FSSH, the quantum wave-packet is propagated along the classical trajectories, and the latter are apportioned among different electronic surfaces according to quantum amplitudes. Hence, the RPSH algorithm carries quantum phase and accounts for electronic coherence.

Recent advancements have been reported for RPSH via propagating the dynamics subject to a generalized Hamiltonian,\cite{Tao:2018} weighted bead-averaged probability of nonadiabatic transition,\cite{Ghosh:2020} and capturing electronic decoherence.\cite{Shakya:2023} In addition, recent refinement schemes have focused on improving sampling strategies,~\cite{Lu:2017,Lu:2018} correct description of electronic coherence,\cite{Kelly2024,Wang2020} and preserving quantum Boltzmann distribution by proper treatment of classically forbidden transitions.\cite{Limbu:2023} 
In the light of the increasing attention to the RPSH simulations and for the purpose of systematically improving the method, here, we evaluate the performance of RPSH in the frameworks of both centroid and bead approximations (RPSH-CA and RPSH-BA, respectively)\cite{Shushkov:2012} in condensed-phase dynamics simulations and investigate the effect of combining them with the diagonal Born-Oppenheimer correction (DBOC).\\ 

DBOC, as the second-order derivative coupling of electronic wavefunctions, is a diagonal term that arbitrarily modifies the BO PESs. It is often neglected in simulations based on mixed quantum-classical dynamics methods due to the absence of a nuclear wavepacket in such schemes.\cite{Tully:2012} Nevertheless, the inclusion of the DBOC in surface hopping methods has been advocated at least for regions with significant nonadiabatic couplings.\cite{Akimov:2013,pssh2009} At the same time, other studies have cautioned about its usage.\cite{dboc2015} It is shown that by adding a repulsive wall in the regions where the two surfaces come close to each other, DBOC deteriorates the results of FSSH simulations.\cite{dboc2016} In comparison, inclusion of NQEs within the RPSH framework, which allows tunneling through an energy barrier, can lead to a different effect than in FSSH. While it would modify the shape of the PESs, the final outcome of the simulations would be dictated by an interrelation between the magnitude of DBOC, the possibility of tunneling via the energy barrier, and the probability of nonadiabatic transitions.

In this work, first, we evaluate the performance of RPSH-CA and RPSH-BA on a generic 2-level system coupled to a classical bath with arbitrary coupling strengths as in the spin-boson model. This model system allows us to probe the capability of the RPSH methods in modeling the population relaxation and preserving quantum Boltzmann distribution in condensed-phase simulations. Then, DBOC will be explicitly incorporated into both frameworks.  In the remainder of this paper, we first give a brief summary of the RPSH-CA and RPSH-BA algorithms as well as methodological considerations regarding improvements of the method. The chosen approach for the inclusion of DBOC in RPSH, as well as the studied model, will also be introduced in the Theory section. Results and Discussions will be presented for dynamics simulations at high and low temperature limits (300 K vs. 30 K) in the context of the spin-boson model, and comparisons will be made to the original FSSH as well as its DBOC-added variant. Concluding Remarks and our Outlook toward future research directions in this field will be presented at the end.

\section{Theory} 
\subsection{\label{sec:RPSH}Ring polymer surface hopping from centroid and bead perspectives}

For a system of $N$ particles in the electronic ground state, the general Hamiltonian is written as:
\begin{equation}
	\hat{H} = \sum_{I=1}^{N} \frac{\hat{\mathbf{P}}_I^2}{2M_I} + \hat{V}(\mathbf{R}), 
	\label{eq:qHam}
\end{equation}
where $\mathbf{R}$ and $\mathbf{P}$ are position and momentum vectors, $M_I$ is the mass of the $I$-th nuclear degree of freedom (DOF), and $ \hat{V}(\mathbf{R})$ is the potential energy surface. The path integral discretization of the quantum mechanical canonical partition function\cite{Feynman:2010,Chandler:1981,Parinello:1984} of this system yields

\begin{equation}
	Z = \mathrm{Tr}\left[ \mathrm{e}^{-\beta\hat{H}} \right]= \lim_{n\to\infty} \frac{1}{(2\pi\hbar)^f}\int d^f\mathbf{R}\int d^f\mathbf{P}\ \mathrm{e}^{-\beta_n H_n(\mathbf{R},
	\mathbf{P})} \,
	\label{eq:qPart}
\end{equation}

where $f=Nn$ and $n$ is the number of imaginary time slices, or ``beads'', of the path integral, and $\beta_n=\beta/n$ with $\beta=(k_\mathrm{B}T)^{-1}$ being the reciprocal temperature. The extended Hamiltonian, $H_n$, is associated with the positions, $\mathbf{R}=\{\mathbf{R}_1,\dots,\mathbf{R}_n\}$, and momenta, $\mathbf{P}=\{\mathbf{P}_1,\dots,\mathbf{P}_n\}$, vectors of the ring polymer and is defined as:   
\begin{equation}
    \begin{split}
	H_n(\mathbf{R},\mathbf{P}) = \sum_{I=1}^N\sum_{j=1}^{n} \left[ \frac{P^2_{I,j}}{2M_{I}} + 
	\frac{1}{2}M_{I}\omega_n^2 \left( R_{I,j}-R_{I,j-1} \right)^2 \right]\\
    + \sum_{j=1}^{n} V\left(R_{1,j},\dots,R_{N,j}\right).
    \end{split}
	\label{eq:piHam}
\end{equation}
The definition of mass in the fictitious kinetic energy term constructs one of the two fundamental differences between running the dynamics with RPMD\cite{Craig:2004} vs. obtaining statistical information from the path integral molecular dynamics (PIMD).\cite{Parinello:1984} The former uses the physical mass of each bead, where the latter may use a fictitious mass. The second term in Eq.~\ref{eq:piHam} represents the harmonic interaction between adjacent beads with $\omega_n=1/\beta_n\hbar$.

Surface hopping approach can take advantage of the extended Hamiltonian in Eq.~\ref{eq:piHam} for propagating the dynamics of the whole ring polymer on a single adiabatic surface $|\alpha;\textbf{R}\rangle$ where the fewest switches ansatz would take it from one surface to another. The only change that is made to the Hamiltonian is introducing the state-dependent potentials as $V_{\alpha}(\textbf{R})=\langle\alpha;\textbf{R}|\hat{V}|\alpha;\textbf{R}\rangle$ giving birth to the RPSH algorithm. Surface-hopping algorithm requires the numerical integration of time-dependent Schr\"{o}dinger equation (TDSE) along the motion of classical trajectories. Originally, two alternative approximations were proposed for this integration in the context of RPSH.\cite{Shushkov:2012} In bead-averaged approximation, i.e., RPSH-BA, TDSE takes the form 
\begin{equation}
i \hbar \dot{c}_{\alpha}(t)=\left[\frac{1}{n}\sum_{j=1}^n V_{\alpha}(\mathbf{R}_j) \right]c_{\alpha}-i\hbar\sum_{\beta}\left[\frac{1}{n} \sum_{j=1}^n \dot{\mathbf{R}}_j\cdot \mathbf{d}_{\alpha\beta}(\mathbf{R}_j)\right]\,c_{\beta}(t),   \label{BA_TDSE} 
\end{equation}
where $c_{\alpha}(t)$ are the time-dependent complex expansion coefficients of the adiabatic basis and $\textbf{d}_{\alpha\beta}(\textbf{R}_j)=\langle\alpha;\textbf{R}_j|\nabla_{\textbf{R}_j}|\beta;\textbf{R}_j\rangle$ is the nonadiabatic coupling vector (NACV) between surfaces with respect to the position of each bead of the ring polymer. 

Alternatively, in the centroid approximation, i.e., RPSH-CA, the position and momentum of the centroid of the ring polymer are updated at every time step of the nuclear dynamics, as:
\begin{equation}
\bar{\textbf{R}}=\frac{1}{n}\sum_{j=1}^n\textbf{R}_j\;\;\;\;\;\;\;\;\;\;\;\;\;    \bar{\textbf{P}}=\frac{1}{n}\sum_{j=1}^n\textbf{P}_j .   
\end{equation}
Then, TDSE is numerically integrated as
\begin{equation}
i \hbar \dot{c}_{\alpha}(t)=V_{\alpha}(\bar{\textbf{R}})\,c_{\alpha}(t)-i\hbar\sum_{\beta}\dot{\bar{\textbf{R}}}\cdot \textbf{d}_{\alpha\beta}(\bar{\textbf{R}})\,c_{\beta}(t)   \label{CA_TDSE} 
\end{equation}
where both the energy of the adiabatic surfaces, $V_{\alpha}(\bar{\textbf{R}})$, and the nonadiabatic coupling vector between surfaces, $\textbf{d}_{\alpha\beta}(\bar{\textbf{R}})=\langle\alpha;\bar{\textbf{R}}|\nabla_{\bar{\textbf{R}}}|\beta;\bar{\textbf{R}}\rangle$, are evaluated at the centroid level. 

The probability of the nonadiabatic transition between surfaces at each time step is defined based on density matrix elements, $\rho_{\alpha\beta}=c_{\alpha}c_{\beta}^{*}$, and the associated derivative coupling. If the transition occurs, the entire ring polymer switches to the new state, and the kinetic energy is accordingly adjusted to conserve the total energy for an RPSH trajectory. 

\subsection{\label{sec:RPSH}Theoretical considerations about the RPSH formalism}

Here, instead of providing the details of the algorithms, which are available elsewhere,\cite{Shushkov:2012,Shakib:2017,Limbu:2023} we discuss and clarify some aspects of RPSH for condensed-phase nonadiabatic simulations:
\begin{itemize} 
    \item During RPSH simulations, we keep the whole ring polymer confined to one electronic state instead of assigning state-specific indices to the beads and spreading the ring polymer over different states.  In fact, the latter has been already investigated in a slightly different flavor of RPSH algorithm.\cite{Lu:2017} While the method was validated numerically, it was inefficient in sampling the off-diagonal elements of observables related to the coupling of different states. This was due to the fact that the major contributors to these elements were configurations with kinks, i.e., consecutive beads residing on different surfaces, and they had a higher energy than configurations without kinks. Allowing an infinite number of switches resulted in a proper sampling of these rare events, but it simply converged the method\cite{Lu:2018} to mean-field ring polymer representation. This is the same observation that led to development of FSSH algorithm where the number of state switches should be minimized to maintain the correct statistical distribution of state populations,\cite{Tully:1990} otherwise the algorithm would regress to a mean-field-like dynamics simulation. Evolving the ring polymer on one surface in RPSH avoids such sampling issues and preserves a surface hopping algorithm suitable for simulating branching and scattering events.
    \item It is accustomed in RPMD simulations that the classical ring polymer Hamiltonian is sampled at an elevated temperature of $\beta_n$ instead of the real temperature $\beta$. This ensures that the positions of all beads are distributed according to the exact quantum Boltzmann distribution.\cite{Craig:2004,Markland:2018} However, multi-state RPSH is different from adiabatic RPMD in that partitioning of trajectories on different electronic states is mainly dictated by the surface hopping algorithm. Using a two-level system coupled to a chain of classical particles, we have shown that RPSH-CA \textit{approximately} reproduces the correct Boltzmann populations subject to a proper treatment of classically forbidden transitions, also known as frustrated hops.\cite{Limbu:2023} 
    This puts the method in a different class than most other multi-state RPMD methods, which fail to reproduce the expected Boltzmann populations, in part due to zero-point energy leakage.\cite{Ananth:2022} Here, we reverse the velocity of frustrated hops along the direction of NACVs at the centroid level to help preserve the quantum Boltzmann distribution. Alternatively, as we have shown before, one can rescale the velocity along the direction of NACVs of each bead of the ring polymers leading to slightly different results. \cite{Shakib:2017} That being said, we remind the readers of our previous studies, where we caution about the possible adverse effect of reversing the velocity of frustrated hops on the quantum dynamics in different model systems.\cite{Limbu:2023} 
   
    \item The RPSH results in the current study are not corrected for the lack of decoherence, or the overcoherence problem of the surface hopping algorithm. A very detailed and comprehensive study on the effect of decoherence correction in simulating the nonadiabatic dynamics of spin-boson model using FSSH method has already been carried out by Chen and Reichman.\cite{Chen:2016} Correcting the non-equilibrium dynamics either with augmented FSSH (A-FSSH) method\cite{Jain:2016} or simply damping the coherence of the density matrix via a pure-dephasing-like rate\cite{Zhu:2004} clearly did very little for improving the results.\cite{Chen:2016} Generally, A-FSSH produces similar population transfer profile to FSSH whereas dephasing damps the oscillatory population decays without necessarily improving the results. On the other hand, other studies have probed the effect of decoherence correction in RPSH and reported promising results for 1D Tully models.\cite{Shakya:2023} However, the lack of a well-defined temperature in such models and the variety of decoherence-correction approaches may require a more comprehensive study in the future.
\end{itemize}

\subsection{\label{sec:dboc}Diagonal Born--Oppenheimer correction}

The action of the nuclear kinetic energy operator, $\hat{T}_\mathrm{n} = -\sum_{I} \frac{\hbar^2}{2 M_I} \nabla_{\mathbf{R}_I}^2$, on the adiabatic electronic wavefunctions results in diagonal and off-diagonal nonadiabatic couplings.~\cite{dboc1997,dboc2016} The former is a potential-like term and will modify the adiabatic PESs, $V_\alpha(\mathbf{R})$, as:
\begin{equation}
\tilde{V}_{\alpha}(\mathbf{R}) = V_\alpha(\mathbf{R}) + V_{\mathrm{DBOC}}^{(\alpha)}(\mathbf{R}).
\end{equation}
The correction term for $N$ ring polymers comprised of $n$ beads is defined as:
\begin{equation}
V_{\mathrm{DBOC}}^{(\alpha)}(\mathbf{R}) = \sum_{I=1}^N\frac{\hbar^2}{2M_{I}} \sum_{j=1}^n \sum_{\beta \neq \alpha} | \mathbf{d}_{I,j}^{(\alpha\beta)}(\mathbf{R}_j) |^2.
\label{eq:vdboc}
\end{equation}

The total force acting on bead $j$ on adiabatic state $\alpha$ is given by:
\begin{equation}
\begin{split}
    \mathbf{F}_{I,j} &= -\nabla\tilde{V_\alpha}(\mathbf{R}_j) = -\nabla V_\alpha(\mathbf{R}_j) - \nabla V_{\mathrm{DBOC}}^{(\alpha)}(\mathbf{R}_j) = \mathbf{F}_{I,j}^{(0)} + \Delta \mathbf{F}_{I,j},
\end{split}
\end{equation}
where $\mathbf{F}_{I,j}^{(0)}$ is the original force without the correction and the DBOC force contribution is given by
\begin{equation}
    \Delta \mathbf{F}_{I,j} =\sum_{\alpha \ne \beta} \frac{1}{M_{I}} \mathbf{d}_{I,j}^{(\alpha \beta)}(\mathbf{R}_j) \cdot \nabla \mathbf{d}_{I,j}^{(\alpha \beta)}(\mathbf{R}_j).
\end{equation}
The NACVs are computed at the individual bead level of the ring polymers according to
\begin{equation}
    \mathbf{d}_{I,j}^{(\alpha \beta)} = \frac{\langle\psi_{j,\alpha}|\nabla_{R_{I,j}} H({\mathbf{R}_j})|\psi_{j,\beta}\rangle}{V_\beta \left({\mathbf{R}_j}\right) - V_\alpha \left({\mathbf{R}_j}\right)}.
\end{equation}

In this work, DBOC forces are computed numerically via finite difference method according to 
\begin{equation}
\nabla \mathbf{d}_{I,j}^{(\alpha \beta)} = \frac{\mathbf{d}_{I,j}^{(\alpha \beta)}(\mathbf{R}_j+\delta) - \mathbf{d}_{I,j}^{(\alpha \beta)}(\mathbf{R}_j-\delta)}{2\delta}.
\end{equation}
The displacement parameter $\delta$ should be chosen carefully to ensure numerical stability and convergence of the DBOC forces within the dynamics simulations.

\subsection{\label{sec:simulation}Simulation details}

The spin-boson model is a well-known system that offers a theoretical framework for examining nonadiabatic dynamics in a condensed phase environment.~\cite{Leggett1987} Details of the model, including discretizations of the bath~\cite{Wang:2001_bath_discret, Hanna:2013_bath_discret}, are provided in the Supporting Information (SI), sections S1-S3 and Fig. S1. Our RPSH schemes investigate the nonadiabatic dynamics of \textit{N} classical bath trajectories, all of which are represented with ring polymers composed of \textit{n} beads. Initial nuclear positions and momenta can be sampled from a normal distribution, 
\begin{equation}
    \rho(\mathbf{R}, \mathbf{P}) = \prod_{I=1}^{N} \prod_{j=1}^{n} \exp\left[ - \beta_n \left(\frac{P^2_{I,j}}{2M_{I}} + \frac{1}{2} M_{I} \omega_{I}^2 R^2_{I,j} \right) \right],
\end{equation}
where $\omega_I$ is the frequency of the $I$-th bath mode. Alternatively, one can employ a normal mode distribution for the ring polymer:
\begin{equation}
    \rho_{\text{RP}}(\tilde{\mathbf{R}}, \tilde{\mathbf{P}}) = \prod_{I=1}^{N} \prod_{k=1}^{n} \exp\left[ - \beta_n \left(\frac{\tilde{P}^2_{I,k}}{2 M_{I}} + \frac{1}{2} M_{I}\omega^2_{I,k} \tilde{R}^2_{I,k} \right) \right],
    \label{eq:int_rpmd}
\end{equation}
where $\omega^2_{I,k} = \omega_I^2 + \omega^2_k$ with $\omega_I$ being the frequency of the $I$-th bath mode and $\omega_k$ the frequency of the $k$-th normal mode. These coordinates can be transformed back to primitive bead coordinates via the inverse normal mode transformation. With proper implementation, these two initial sampling methods should yield similar results. Here, the initial nuclear positions and momenta are sampled from a normal mode distribution consistent with the ring polymer representation as in Eq.~\ref{eq:int_rpmd}. 

All simulations are carried out using 10,000 independent trajectories, each coupled to a Debye spectral bath~\cite{Sun:1998,Richardson:2019} consisting of 100 harmonic oscillator modes. The bath frequencies are sampled according to a Debye spectral density~\cite{Miller1999-debye,Miller2001-debye,Templaar:2018} with appropriate reorganization energy~\cite{Ishizaki2009_Er, Cotton:2016} and cutoff frequency. Trajectories are propagated for 0.5 ps using a time step of 0.0242 fs (1 a.u.). The initial electronic wave packet is prepared and projected onto the adiabatic basis, and the initial active surface, $\lambda$, is selected based on initial electronic amplitudes, i.e., trajectories are placed on surface $\alpha$ with probability $|c_\alpha|^2$. Diabatic populations are computed on-the-fly using the mixed quantum-classical density matrix formalism as described in Ref.~\citenum{Subotnik2013:populaiton}:
\begin{equation}
P_\text{\tiny{D}} = \left\langle \sum_\alpha |U_{\text{\tiny{D}} \alpha}|^2 \delta_{\alpha,\lambda} + \sum_{\alpha<\beta} 2\, \text{Re}\left[ U_{\text{\tiny{D}} \alpha} , c_{\alpha}c_{\beta}^{*} , U_{\text{\tiny{D}} \beta}^* \right] \right\rangle,
\label{eq:pop}
\end{equation}
where $U_{\text{\tiny{D}}\alpha}$ are elements of the adiabatic-to-diabatic transformation matrix, $\lambda$ denotes the active surface, and $\rho_{\alpha\beta}=c_{\alpha}c_{\beta}^{*}$ is the electronic density matrix. Final populations are obtained by averaging over the full ensemble of trajectories.

\begin{figure*}
    \centering
    \includegraphics[width=.99\linewidth]{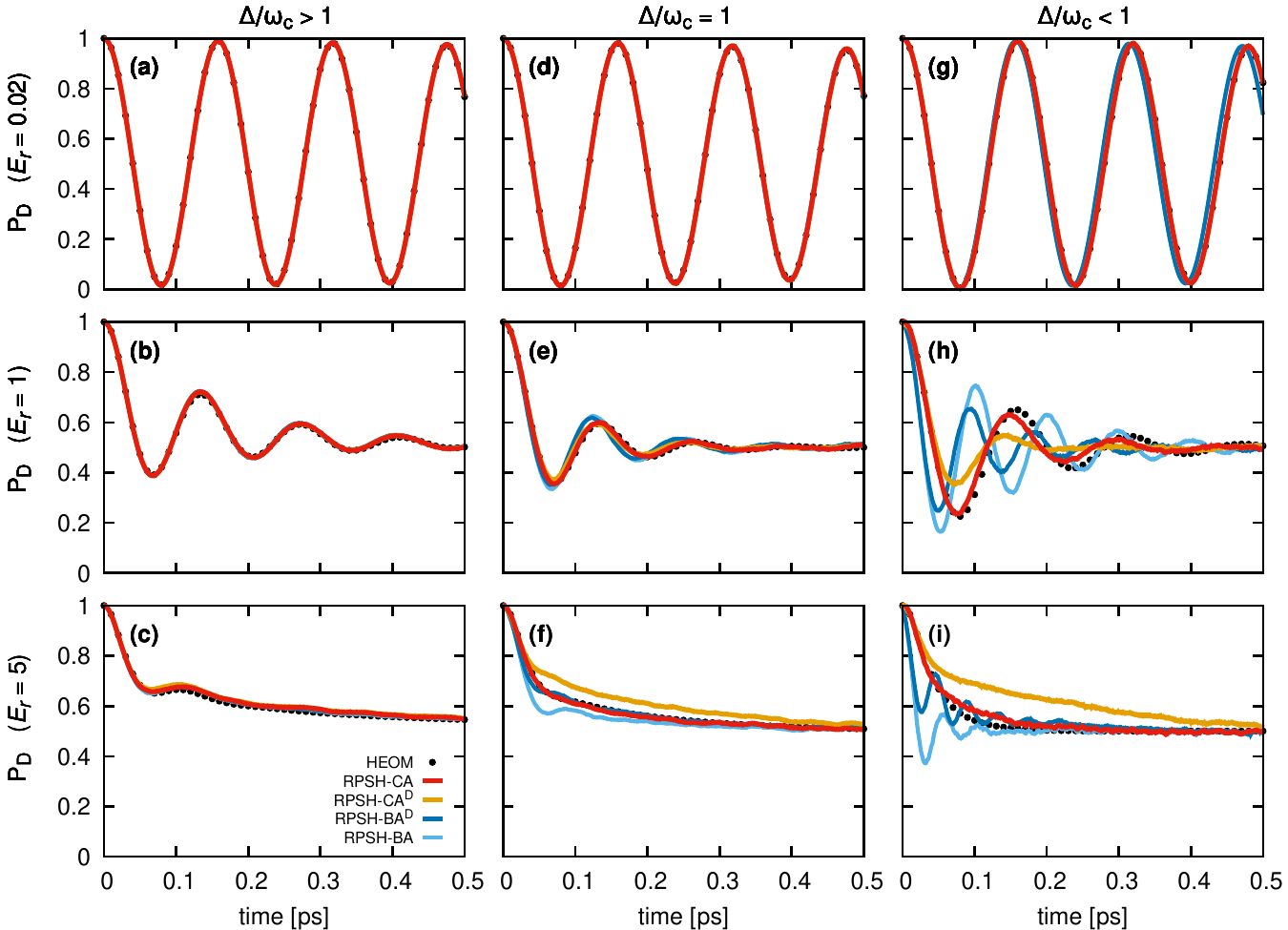}
    \caption{Population dynamics for the spin-boson model at the high temperature of 300 K and three reorganization energies, $E_r$ = 0.02 (top row), $E_r$ = 1 (middle row), and $E_r$ = 5 (bottom row), across different reaction regimes (columns). Model parameters include $\epsilon$ = 0, $\Delta$ = 1.0, and $\omega_c$ = \{0.25$\Delta$, $\Delta$, 5$\Delta$\}. All energies are in the unit of 104.25 cm$^{-1}$.}
    \label{fig:pop-dboc-highT}
\end{figure*}

\section{Results and Discussion} 

We systematically evaluate the applicability of RPSH in condensed-phase dynamics simulations by investigating population dynamics of first symmetric and then asymmetric spin-boson models across various temperatures, reorganization energies, $E_r$, and reaction regimes. At the same time, we assess the effect of incorporating DBOC within the RPSH framework. Exact quantum results based on the hierarchical equations of motion (HEOM)~\cite{heom,qutipheom} are computed using QuTiP~\cite{qutip} and serve as benchmarks throughout. All the RPSH simulations are conducted using 32 beads, including RPSH-CA and RPSH-BA, as well as their new variants with DBOC, namely RPSH-CA$^\text{D}$ and RPSH-BA$^\text{D}$, respectively, with the superscript D denoting the inclusion of DBOC terms. Convergence of results with respect to the number of beads is shown for intermediate and nonadiabatic reaction regimes and low-temperature limits in the SI, Fig. S2 and S3. Note that this number of beads is only followed for the sake of consistency in the current paper. For most reaction regimes, such as the adiabatic reaction regime at high or low temperature,  the results are already converged with only four beads. All the simulations are carried out up to 0.5 ps. By definition, the original RPMD method is expected to produce exact quantum dynamical properties at the limit that time goes to zero ($t\rightarrow0$),
albeit as $n\rightarrow\infty$.\cite{Craig:2004} In the current work, we refer to the time limit closer to zero as short-time and the time limit closer to 0.5 ps as long-time limits. With this distinction, we investigate how the RPSH method fares moving away from the $t\rightarrow0$ limit.

\begin{figure*}[t]
    \centering
    \includegraphics[width=.99\linewidth]{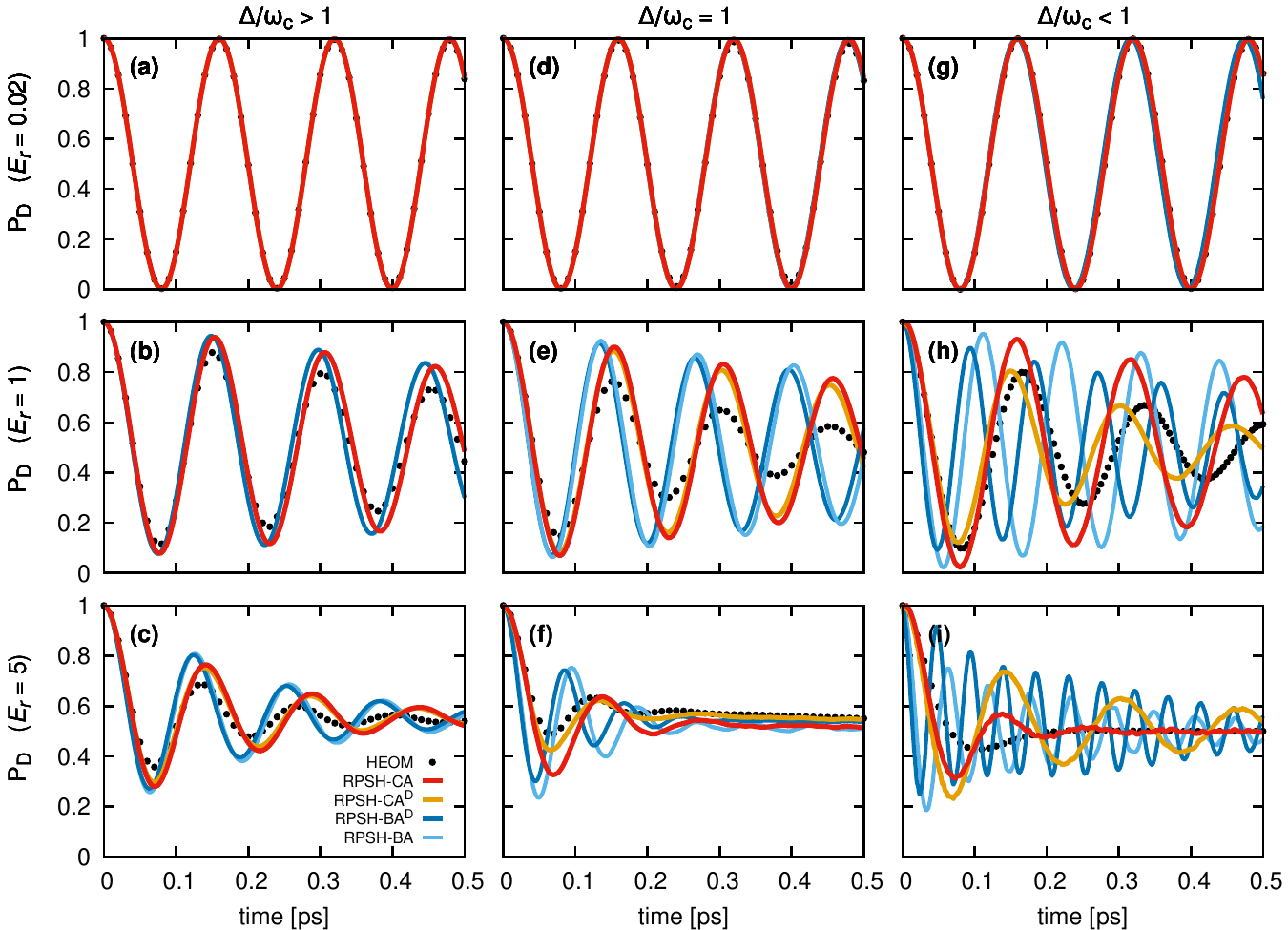}
    \caption{Population dynamics of spin-boson model at the low temperature of 30 K for three reorganization energies and different reaction regimes. All the model parameters are the same as in Fig.~\ref{fig:pop-dboc-highT}.}
    \label{fig:pop-dboc-lowT}
\end{figure*}

\subsection{Symmetric spin-boson nonadiabatic dynamics}
Fig.~\ref{fig:pop-dboc-highT} shows the change of population of the donor diabatic electronic state ($P_D$) at a high temperature of 300 K. The population dynamics of the spin-boson model is reported for three reorganization energies: $E_r=0.02$, $E_r=1$, and $E_r=5$, each examined in adiabatic ($\Delta/\omega_c > 1$), intermediate ($\Delta/\omega_c = 1$), and nonadiabatic ($\Delta/\omega_c < 1$) reaction regimes. These reaction regimes correspond to slow, comparable, and fast motions of baths relative to the electronic coupling, $\Delta$, respectively.

As shown in Fig.~\ref{fig:pop-dboc-highT}(a), there is an excellent agreement between the simulated population dynamics using four RPSH variants and the exact quantum results in the adiabatic reaction regime. Coherence, oscillation amplitude, and phase align closely with HEOM benchmarks. As the coupling of the quantum sub-system to the classical solvent increases, Fig.~\ref{fig:pop-dboc-highT}(a)-(c), all four methods keep performing very well in reproducing exact results. 

Fig.~\ref{fig:pop-dboc-highT}(d)-(f) demonstrate the population transfer profile in the intermediate reaction regime where all four methods are exact at $E_r=0.02$. As the reorganization energy increases, RPSH-CA remains very reliable, producing quantum populations close to exact results. RPSH-BA starts showing small deviations from exact results, more apparent at the highest degree of coupling to the bath, Fig.~\ref{fig:pop-dboc-highT}(f), while RPSH-BA$^\text{D}$ improves the performance to a great deal. On the other hand, DBOC inclusion has a deteriorating effect on the performance of RPSH-CA in this regime, compare yellow vs. red line in Fig.~\ref{fig:pop-dboc-highT}(f).

Fig.~\ref{fig:pop-dboc-highT}(g)-(i) demonstrate the population transfer profile in the nonadiabatic reaction regime. Here, RPSH-BA shows deviations from the exact results at all values of $E_r$, while RPSH-CA performs satisfactorily well in reproducing the correct phase oscillation in the short-time limit and the Boltzmann populations at the long-time limit as the $E_r$ increases. The diagonal correction adds a repulsive wall to the ground state adiabatic PES when the energy of the two states gets close, thus hindering the population transfer from the donor to the acceptor side. This leads to a deteriorating effect on $P_D$ profile in RPSH-CA$^\text{D}$. However, the same change of PES has an improving effect on RPSH-BA, which was already suffering from an exaggerated population transfer from reactant to product. As a result, in comparison to RPSH-BA, RPSH-BA$^\text{D}$ produces better agreement with the exact results in terms of the amplitude of population oscillation but is still unreliable in terms of correct phase oscillations.

\begin{figure*}
    \centering
    \includegraphics[width=.99\linewidth]{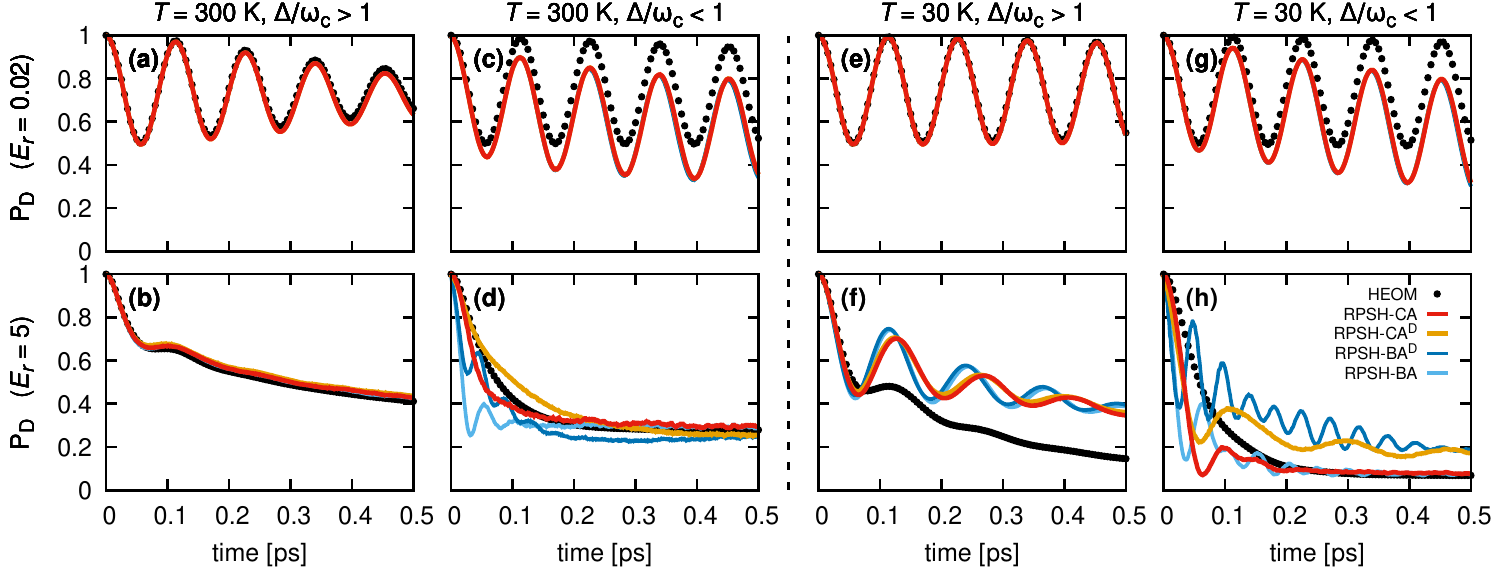}
    \caption{Population dynamics of the asymmetric spin–boson model at 300 K (a-d, left panel) and 30 K (e-h, right panel) for low and high reorganization energies: $E_r = 0.02$, top panel, and $E_r = 5$, bottom panel, spanning both adiabatic and nonadiabatic reaction regimes. Results from RPSH-CA and RPSH-BA and their DBOC variants (RPSH-CA$^{\text{D}}$ and RPSH-BA$^{\text{D}}$) are compared against exact quantum dynamics results. Here, $\epsilon$ = 1.0 and all other model parameters are the same as in Fig.~\ref{fig:pop-dboc-highT}.}
    \label{fig:pop-dboc-bias}
\end{figure*}

Fig.~\ref{fig:pop-dboc-lowT} demonstrates the population transfer profile of the spin-boson model at different reaction regimes at a low temperature of 30 K. The population dynamics are presented for the same set of reorganization energies: $E_r=0.02$, $E_r=1$, and $E_r=5$, analyzed under slow, intermediate, and fast bath motions. Here, the adiabatic reaction regime with the lowest $E_r$ of 0.02, Fig.~\ref{fig:pop-dboc-lowT}(a),~\ref{fig:pop-dboc-lowT}(d), and~\ref{fig:pop-dboc-lowT}(g), is where all the four methods work very well in accordance with exact simulations. Increasing the reorganization energy to $E_r=1$ in Fig.~\ref{fig:pop-dboc-lowT}(b), RPSH-CA and RPSH-CA$^\text{D}$ preserve the correct Rabi oscillations in the 0.5 ps time frame while the amplitudes of the population oscillations becomes higher than the exact results as the time goes on. On the other hand, RPSH-BA and RPSH-BA$^\text{D}$ show deviation from both amplitude and phase of oscillations as time goes on. Nevertheless, all methods work satisfactorily in this regimes at the short-time limit. At the highest reorganization energy of $E_r=5$, Fig.~\ref{fig:pop-dboc-lowT}c, all methods show deviations from correct oscillations. 

The effect of inclusion of DBOC in RPSH is more apparent in intermediate and nonadiabatic reaction regimes with higher reorganization energies of $E_r=1$, Fig.~\ref{fig:pop-dboc-lowT}(e) and~\ref{fig:pop-dboc-lowT}(h), and $E_r=5$, Fig.~\ref{fig:pop-dboc-lowT}(f) and~\ref{fig:pop-dboc-lowT}(i). However, first we need to discuss the elephant in the room which is the nonphysical oscillatory behavior of RPSH-BA and RPSH-BA$^{\text{D}}$ in these reaction regimes. A comparison of the nonadiabatic transition statistics over 10000 RPSH-CA and RPSH-BA trajectories shows a similar frequency of successful or frustrated hops in almost all reaction regimes in both high and low temperature, see Fig. S4 and S5 in the SI. The slight differences of the frequencies of hops in the high reorganization energy and nonadiabatic reaction regime cannot explain the very clear difference in the behavior of RPSH-BA vs. RPSH-CA seen here. Note that the diabatic donor populations are computed using
the mixed quantum-classical density matrix formalism in Eq.~\ref{eq:pop} which in turn is dominated by the time evolution of the electronic state coefficients via Eq.~\ref{BA_TDSE} in RPSH-BA and Eq.~\ref{CA_TDSE} in RPSH-CA. Fig. S6(a)-S6(c) in the SI show the time evolution of $c_{\alpha}$ for the same trajectory in RPSH-BA vs. RPSH-CA. As can be seen, the former shows very intense oscillations compared to the smooth behavior of the latter. Taking the averages over the quantities of each of the 32 beads of all 100 classical ring polymers, randomly distributed over the electronic PESs, leads to this oscillatory behavior whereas the narrower distribution of the centroids of ring polymers leads to a smooth behavior free from nonphysical oscillations. It should be mentioned that this difference between RPSH-BA and RPSH-CA is seen due to the high number of classical DOFs and the disparate spatial distributions of different ring polymers. Such an observation would not be apparent in the earlier studies of RPSH which involved only one classical DOF.\cite{Shushkov:2012} This issue can be shown even here by reducing the number of beads of ring polymers from 32 to 4, Fig. S6(d)-S6(f) in the SI, which leads to more tamed oscillations of the electronic coefficients over time in RPSH-BA while it does not have such a pronounced effect on RPSH-CA.

Based on these observations, here, we will only discuss the effect of DBOC on RPSH-CA while cautioning the community about using RPSH-BA for inclusion of NQEs into the condensed-phase MD simulations at low temperature limits in spite of the early reports on its success\cite{Shushkov:2012} in 1D scattering models. At the highest reorganization energy of the intermediate reaction regime, Fig.~\ref{fig:pop-dboc-lowT}(f), RPSH-CA$^{\text{D}}$ acts more reliably in capturing the correct  population profile in comparison to the original RPSH-CA. In $E_r=1$ in the nonadiabatic reaction regime, Fig.~\ref{fig:pop-dboc-lowT}(h), the dampened dynamics of RPSH-CA$^\text{D}$ due to the repulsive wall on the PES corrects the overestimation of population transfer happening in RPSH-CA. As a result, RPSH-CA$^\text{D}$ reproduces the correct population amplitude, albeit with shifted oscillations. Fig.~\ref{fig:pop-dboc-lowT}(i) shows the population transfer profile at $E_r=5$ in the nonadiabatic reaction regime. Here, RPSH-CA is very successful in reproducing the quantum population in the long-time limit. However, it faces nonphysical oscillations in the short-time limit. These oscillations, which become more apparent in RPSH-CA$^\text{D}$, can be the result of internal high-frequency oscillations of the ring polymers affected by the number of beads. Fig. S2(d) in the SI shows that using 4 beads instead of 32 beads would smooth out such oscillations in RPSH-CA, still preserving correct population in the long-time limit.

\subsection{Asymmetric Spin-boson nonadiabatic dynamics}

The asymmetric spin–boson model with a Debye bath presents a more stringent benchmark for any approximate quantum dynamics method than the symmetric case due to the inherent energy bias between diabatic states.~\cite{Runeson2019_JCP,Liu2024_JCP,Amati2023_JCP} On a related note, the original RPMD has an inferior performance in an anhormonic potential than the harmonic potential.\cite{Craig:2004} Hence, it is interesting to see the performance of RPSH in challenging cases such as asymmetric spin-boson model. Fig.~\ref{fig:pop-dboc-bias} compares the performance of RPSH-CA, RPSH-CA$^{\text{D}}$, RPSH-BA, and RPSH-BA$^{\text{D}}$ against exact HEOM dynamics at 300 K and 30 K for $\epsilon = 1$, $\Delta = 1$, and under the lowest and highest reorganization energies of $E_r = 0.02$, and 5. The results cover adiabatic and nonadiabatic reaction regimes with $\omega_c = 0.25$, and 5. At high temperature and slow bath motion, all methods reproduce the overall relaxation and equilibrium populations with good accuracy at both $E_r = 0.02$, and 5, Fig.~\ref{fig:pop-dboc-bias}(a) and~\ref{fig:pop-dboc-bias}(b). However, stark deviations occur at both cases with the fast bath motions. With $E_r = 0.02$, all four methods similarly deviate from the exact behavior, Fig.~\ref{fig:pop-dboc-bias}(c). With $E_r = 5$, while they perform poorly at the shot-time limit, they recover toward the long-time limit. At the low temperature limit, we notice  poorer performance from RPSH. Fig.~\ref{fig:pop-dboc-bias}(e) presents an exact performance by all methods at the adiabatic reaction regime and the lowest coupling to the environment. However, increasing $E_r$ to 5 all methods completely fail in reproducing the correct population transfer profile. Their performance is not much better in the nonadiabatic reaction regime with the exception of RPSH-CA and, surprisingly, RPSH-BA being able to retrieve correct Boltzmann population at $E_r=5$, Fig.~\ref{fig:pop-dboc-bias}(h).  Overall, in the studied asymmetric models, the DBOC inclusion provided no improvements in the obtained results.\\

\subsection{DBOC effect in RPSH versus FSSH}
\begin{figure}
    \centering
    \includegraphics[width=.99\linewidth]{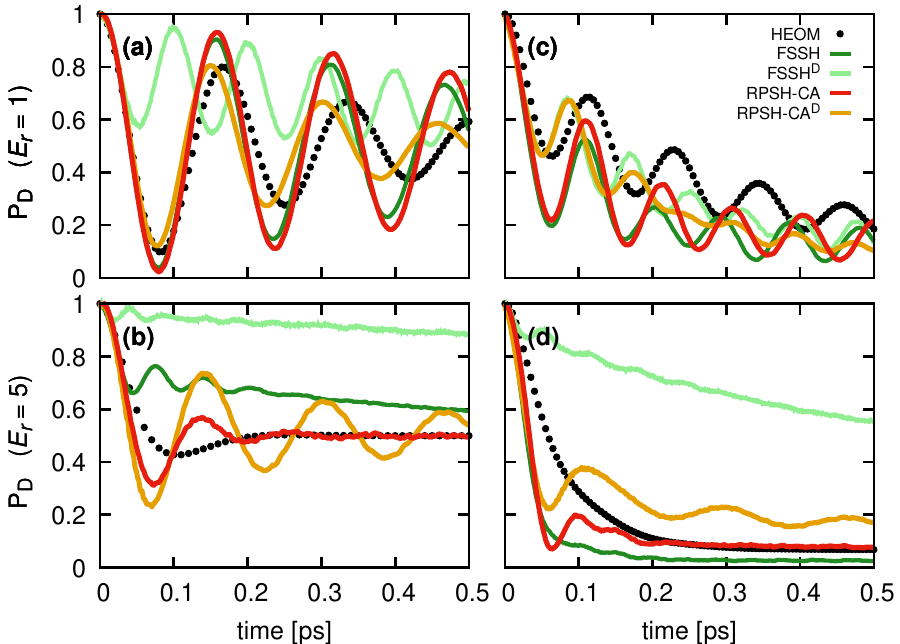}
    \caption{Population dynamics of the symmetric (a-b) and asymmetric (c-d) spin-boson models at a low temperature of 30 K, shown for different reorganization energies in the nonadiabatic reaction regime. The plots compare results from FSSH, RPSH, and their DBOC-corrected variants (FSSH$^{\text{D}}$ and RPSH-CA$^{\text{D}}$) against exact results.}
    \label{fig:pop-fssh-rpsh}
\end{figure}

Focusing on the nonadiabatic reaction regime, with high reorganization energies of 1 and 5, Fig.~\ref{fig:pop-fssh-rpsh} compares and contrasts the performance of RPSH-CA and RPSH-CA$^{\text{D}}$ to the classical FSSH and its DBOC variant (FSSH$^{\text{D}}$) at low temperature of 30 K. The comparison is made in both symmetric, Fig.~\ref{fig:pop-fssh-rpsh}(a) and~\ref{fig:pop-fssh-rpsh}(b), and asymmetric, Fig.~\ref{fig:pop-fssh-rpsh}(c) and~\ref{fig:pop-fssh-rpsh}(d), cases. As can be seen in Fig.~\ref{fig:pop-fssh-rpsh}(a) and~\ref{fig:pop-fssh-rpsh}(c), both RPSH-CA and FSSH perform poorly in reproducing the population transfer profile at $E_r=1$. However, in symmetric potential, RPSH-CA$^{\text{D}}$ recovers the correct amplitude of population oscillations if not the phase. At the same region, DBOC has a completely deteriorating effect on FSSH results. It can inferred that, in accordance with the literature,\cite{dboc2015,dboc2016} the DBOC repulsive wall on PES at the barrier crossing hinders the population transfer between the donor and the acceptor sites in the FSSH$^{\text{D}}$ simulations. However, RPSH-CA$^{\text{D}}$ encounters a more smooth change of the shape of the PES due to the distribution of ring polymers which does not hinder the population transfer as in the case of FSSH. In the asymmetric case, while DBOC seems to shift both RPSH-CA and FSSH populations towards the exact results but it does not leave a lasting improving effect on either method.

As the reorganization energy increases to $E_r=5$ in Fig.~\ref{fig:pop-fssh-rpsh}(b) and~\ref{fig:pop-fssh-rpsh}(d), FSSH fails to reproduce the correct population transfer profile while DBOC variant has even more devastating effect on the results. Obviously the repulsive wall on the PES completely dampens the population transfer between the donor and acceptor sites. On the other hand, in both symmetric and asymmetric case, RPSH-CA is capable of reproducing the long-time Boltzmann population even though showing nonphysical oscillations in the short-time limit. Overall, the deteriorating effect of DBOC on the RPSH-CA results is not as extreme as the FSSH case. One might even argue that the nonphysical oscillations in RPSH-CA$^{\text{D}}$ can be dampened with a proper decoherence scheme leading to more improved results.

To be more precise about the effect of DBOC in RPSH vs. FSSH, we repeat that the height of the DBOC repulsive wall on adiabatic PESs in RPSH-CA$^\text{D}$ cannot be as high as the case in FSSH$^\text{D}$. The DBOC in RPSH-CA$^\text{D}$ is calculated based on Eq.~\ref{eq:vdboc}, which takes the average of NACVs of all beads of the ring polymers. Since the corresponding NACVs of different beads spike at slightly different positions, taking the average leads to a smoother change of the shape of PESs compared to one NACV in a classical FSSH trajectory. This point is more clarified in a test-case study where we calculate DBOC for RPSH-CA$^\text{D}$ at the centroid level instead of every bead of the ring polymers. Since the centroid is like one individual classical particle, the DBO correction of RPSH-CA at the centroid level leads to a very steep repulsive potential wall and basically similar results to FSSH$^{\text{D}}$, see Fig. S7 in the SI. Overall, it can be said that the RPSH-CA$^{\text{D}}$ method either improves the RPSH-CA results or, at least, does not diminish them to the degree that happens in the FSSH method. Therefore, we suggest DBOC as a parameter that worth to be tested when dealing with different systems of interest the results of which, of course, are method-dependent.

\section{Conclusions and Outlooks}
In this study, we developed and tested different RPSH methods that incorporate the diagonal Born–Oppenheimer correction alongside derivative couplings into nonadiabatic MD simulations, focusing on the spin-boson model as a benchmark. We employed two different versions of RPSH, where nonadiabatic transitions are captured at the centroid or bead level, RPSH-CA and RPSH-BA, respectively. By comparing these methods and the standard FSSH to their DBOC-included variants across various reorganization energies and reaction regimes, we systematically explored the influence of geometric corrections on quantum dynamics. Our results, overall, demonstrate the utility of RPSH methods for condensed-phase MD simulations at physically meaningful temperatures. It is a step forward in the utilization of RPSH in comparison to 1D model systems with arbitrary temperatures employed previously. Across most reaction regimes examined in a symmetric potential, RPSH-CA consistently preserves quantum coherence and maintains satisfactory oscillation amplitude and phase, even under strong system–bath coupling strengths. In some reaction regimes, the addition of DBOC can further improve the accuracy of RPSH-CA results. On the other hand, RPSH-BA and its DBOC variant deviate considerably from correct dynamics behavior at the low temperature of 30 K unless in the adiabatic reaction regime or very weak coupling of the quantum sub-system to the classical environment. In comparison between RPSH-CA and the conventional FSSH, inclusion of DBOC shows progressive deterioration in performance of the latter at reaction regimes corresponding to fast bath motions or as the reorganization energy increases.

Overall, considering the potential positive effect of including DBOC in RPSH-CA simulations, albeit with the observed system-dependent behavior, it as a parameter worth to be tested when dealing with different problems of interest. Therefore, these new developments are available for community use within the newest version of our user-friendly and highly parallelized software package, SHARP pack.\cite{Limbu:2024} Future research should be directed towards systematic improvement of the accuracy of RPSH simulations for both symmetric and asymmetric potentials. Improving the efficiency of the method using recently introduced techniques that allow convergence of real-time path integral simulations  with a fraction of the number of beads\cite{bfcmd} is another exciting research direction. Finally, with our interest in simulating charge transport dynamics in functional layered materials,\cite{acsami_13_25270,acsami_15_9494,acsaem_7_5143,jpcc_129_2222,jcp_156_044109} we foresee an interface of the SHARP pack with our highly-parallelized MD software package DL\_POLY Quantum.\cite{aprLondon2024,London2025}
\section*{Supplementary material}
The supplementary material contains details of the spin-boson model Hamiltonian, the discretization of the Debye spectral density, RPSH bead convergence tests of population and temperature, nonadiabatic hopping statistics, and the comparison of the population dynamics by FSSH, FSSH$^{\text{D}}$, and two different DBOC variants of RPSH-CA at the low temperature of 30 K.

\begin{acknowledgments}
This research used resources from Bridges2 at Pittsburgh Supercomputing Center through allocation  PHY240170 and CHE200007 from the Advanced Cyberinfrastructure Coordination Ecosystem: Services \& Support (ACCESS) program,\cite{access} which is supported by U.S. National Science Foundation grants \#2138259, \#2138286, \#2138307, \#2137603, and \#2138296. The use of computing resources and support provided by the HPC center at NJIT is also gratefully acknowledged.
\end{acknowledgments}

\section*{Conflict of Interest}
The authors have no conflicts to disclose.

\section*{Data Availability Statement}
The data that support the findings of this study are available within the article and its supplementary material. All data presented in this work were generated using the SHARP Pack package, which is available for download at https://github.com/fashakib/SHARP\_pack\_2. The SHARP Pack documentation is also available on https://sharppack.readthedocs.io/en/latest/. 

%\nocite{*}
\section*{REFERENCES}
\bibliography{bib}

\end{document}